\documentclass[preprint, 11pt, authoryear]{elsarticle}
\pdfoutput=1 

\usepackage{geometry}
 \usepackage{graphicx,booktabs,calc}
 
 \usepackage{pifont}

\usepackage{lineno,hyperref}
\usepackage{hyperref}
\usepackage{multirow}
\usepackage{amssymb}
\usepackage{subcaption} 
\usepackage{xcolor}
\usepackage[colorinlistoftodos,prependcaption]{todonotes}
\usepackage[flushleft]{threeparttable}
\usepackage{enumitem}
\usepackage{makecell}
\usepackage{array}
\usepackage{adjustbox}
\usepackage{rotating}
\usepackage{amsmath,mathtools}
\usepackage{pdflscape}
\usepackage[font=small,labelfont=normalsize]{caption}
\usepackage{lineno}
\journal{arXiv.org}

\biboptions{authoryear}

\begin{document}

\begin{frontmatter}

\title{Spatiotemporal Characteristics of Ride-sourcing Operation in Urban Area}

\author[smart]{Simon Oh}
\author[smart]{Daniel Kondor}
\author[smart]{Ravi Seshadri}
\author[smart]{Meng Zhou}
\author[smart]{Diem-Trinh Le}
\author[mitcee]{Moshe Ben-Akiva}
\address[smart]{Future Urban Mobility, Singapore-MIT Alliance for Research and Technology (SMART)
1 CREATE Way, \#09-02 CREATE Tower, Singapore 138602}
\address[mitcee]{Department of Civil and Environmental Engineering, Massachusetts Institute of Technology, Cambridge, MA 02139, United States}



%

\begin{abstract}
The emergence of ride-sourcing platforms has brought an innovative alternative in transportation, radically changed travel behaviors, and suggested new directions for transportation planners and operators. This paper provides an exploratory analysis on the operations of a ride-sourcing service using large-scale data on service performance. Observations over multiple days in Singapore suggest reproducible demand patterns and provide empirical estimates of fleet operations over time and space. During peak periods, we observe significant increases in the service rate along with surge price multipliers. We perform an in-depth analysis of fleet utilization rates and are able to explain daily patterns based on drivers' behavior by involving the number of shifts, shift duration, and shift start and end time choices. We also evaluate metrics of user experience, namely waiting and travel time distribution, and explain our empirical findings with distance metrics from driver trajectory analysis and congestion patterns. Our results of empirical observations on actual service in Singapore can help to understand the spatiotemporal characteristics of ride-sourcing services and provide important insights for transportation planning and operations.
\end{abstract}


\end{frontmatter}


\section{Introduction}\label{sec:sec_1_intro}
During the last decade, the ride-sourcing market has significantly grown, reaching 7--8.3\% of worldwide penetration rate and generating 32--44 billion USD of revenue in 2016 and 2017 (\cite{oecdreport2018}); this trend is likely to continue, with the penetration rate is projected to reach 13\% in the next five years (\cite{statistareport2018}). The huge success of the ride-sourcing business can be attributed to the wide availability of smartphones with internet service, ease-of-use and on-line payment system, while providing a ride service with a `reasonable’ price and waiting time (\cite{rayle2016just},  \cite{sfmtareport2018}). 

A key aspect of ride-sourcing platforms is to connect two groups from demand and supply side: user (i.e. passengers) and service providers (i.e. driver-partners). Once passengers send their pick-up request through a smartphone app with travel details including origin, destination, and a chosen mode from a given menu (e.g. options for single or shared ride), then `idled’ drivers (who are on-line on the platform, but cruise or wait for their next service assignment) will have a chance to decide to participate or not based on their expectations of utility. The decision to participate is done on different timescales however. Passengers can decide for each trip whether to use a taxi or ride-sourcing service, use an alternative form of transportation (e.g. public transit), or not take the trip at all. This decision can be based on multiple factors, including their current value of time, the utility (or importance) of the trip, and the estimated price and waiting time.

Modeling this two-sided market becomes an important challenge when thinking of current technological trends in urban transportation. Also, fleet management is known as one of the key components that determines the quality of service. Several studies have utilized simulation frameworks to address this operational problem involving vehicle assignment and rebalancing. \cite{santi2014quantifying} implemented shareability networks and shown the possible reduction of taxi fleets and total trip length with ride-sharing in a simulation based on New York taxi data. Similarly, \cite{vazifeh2018addressing} addressed the minimum fleet problem in ride-sourcing using a graph-theoretic method, showing reductions in fleet size compared with current taxi operations. Also, simulation methods have been widely used in other studies to solve assignment and rebalancing problems (\cite{zhang2016control}, \cite{alonso2017demand}, \cite{hyland2018dynamic}). In addition, the performance of on-demand vehicles has been tested in agent-based simulations. \cite{fagnant2014travel} evaluated the impact of shared automated vehicles with simulations in a grid network representing Austin, Texas, and suggested that an on-demand service may help to reduce private vehicle use and the need for parking facilities. \cite{boesch2016autonomous} showed a non-linear relation between the required number of service vehicles and satisfied requests in Zurich. Later, \cite{horl2019fleet}  improved the performance of ride-sharing with different fleet sizes and rebalancing algorithms. Similar conclusion can be found in a simulation study in Berlin (\cite{bischoff2016simulation}) that solves the dynamic vehicle routing problem. More recently, in a case study focusing on Greenwich, London, \cite{segui2019simulating} have assessed different operational settings of ride-sharing services and evaluated multiple strategies from the perspective of traveller, operator, and city. Furthermore, the future impacts of automated ride-sharing system (called automated mobility-on-demand, AMOD) have been evaluated with an activity- and agent-based simulation in Singapore (\cite{oh2020evaluating}), \cite{oh2020amod_tra}. 

Some empirical studies have been conducted to understand the efficiency of a ride-sourcing system. \cite{cramer2016disruptive} analyzed data provided by Uber in the Los Angeles and Seattle areas, and showed higher capacity utilization than traditional taxi services. A large body of research can be found exploring the effect of surge pricing (\cite{hall2015effects}, \cite{battifarano2019predicting}), typically used in ride-sourcing to satisfy more ride requests under the inequality of demand and supply (i.e.~to attract more drivers during the peak period). In the case of Uber in major cities of the United States, it was reported that 8\% to 28\% of the time was operated with surging (\cite{cohen2016using}, \cite{chen2015dynamic}, \cite{battifarano2019predicting}). This surge multiplier has been analyzed from a theoretical economic (\cite{guda2017strategic}, \cite{yang2011equilibrium}, \cite{zha2016economic}, \cite{guha2018dynamics}) and equilibrium model perspective (\cite{wang2016pricing}, \cite{bimpikis2019spatial}, \cite{zha2016economic}, \cite{sun2019model}). More recently, predictive models have been utilized to predict the short-term changes in surge pricing (\cite{ke2017short}, \cite{battifarano2019predicting}).

However, most of the previous studies have been performed with assumptions on operational aspects of ride-sourcing (i.e. pricing, fleet sizes, shift behaviors) that have not been properly calibrated with actual observations (i.e. waiting time, travel cost). In addition, key operational metrics from users' and service providers' standpoint have not been explored in detail to the best of our knowledge. Motivated by these research gaps, this paper conducts an exploratory analysis on the actual operations of a ride-sourcing service. The objective of this work can be summarized as follows:
\begin{enumerate}[label=(\roman*)]
    \item Understanding the operational characteristics of ride-sourcing with the distributions of system measures on (a) service metrics (i.e. service rates, delay, queue, fleet operational details) and (b) user experience (i.e. waiting and travel times) that can be useful for transportation operating and management authorities.
    \item Establishing good empirical estimates (on fleet operations, shift behavior of drivers, service rate and system capacity) that are required in designing future scenarios pertaining to on-demand shared mobility systems.
    \item And, estimating the zone-to-zone travel cost (i.e. waiting and travel time) to generate a travel \textit{skim} which is crucial for the reliability of demand models and useful for transportation planners.
\end{enumerate}

The remainder of this paper is organized as follows. Section \ref{sec:sec_2_ridesourcingsystem} explains the ride-sourcing system with key players (passenger/traveler, driver-partner, service platform) and their interactions. Then, we describe the ride-sourcing data and measurements in Section \ref{sec:sec_3_data}, followed by, in Section \ref{sec:sec_4_empirical_analysis}, an analysis of the spatio-temporal distributions of measurements from two categories: (a) Service metrics and (b) User experience. In this section, we also discuss the potential application of our empirical estimates. Finally, in Section \ref{sec:sec_5_conclusion}, we summarize the key findings and conclude the paper with future research directions.

\section{Ride-sourcing System}\label{sec:sec_2_ridesourcingsystem}
In general, the objectives of ride-sourcing systems is to balance the demand (ride requests) and supply (the available service fleet) over different periods of time and space. In this way, the platform aims to maximize the ``matching rate'' to increase (short-term/long-term) revenue, consumer surplus, and social welfare as well as market penetration. To achieve this goal, ride-sourcing platforms employ operational strategies such as (i) dynamic pricing that charges a different service fare with surge multiplier to encourage the drivers' participation, (ii) assignment algorithm that efficiently matches traveler's request with service vehicles, and (iii) routing guidance that helps drivers to avoid congested links and reduce the wait and travel time of travelers. Further comprehensive review on the framework of ride-sourcing system is available in \cite{wang2019ridesourcing}.

\begin{figure}[!ht]
  \centering
  \includegraphics[width=0.7\textwidth]{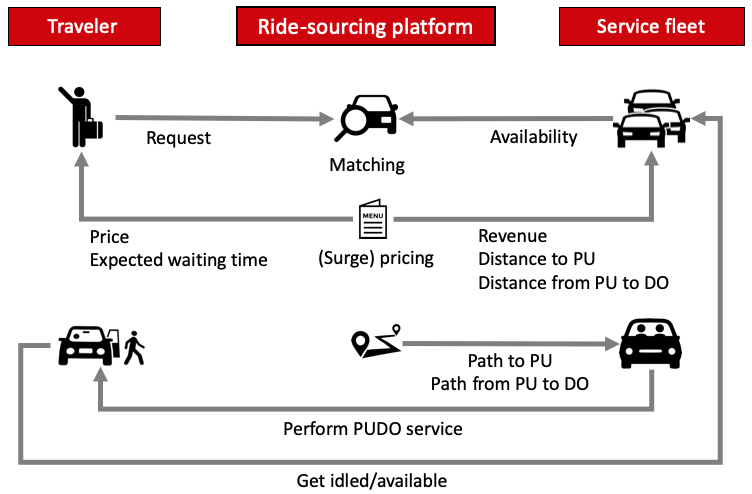}
  \caption{Schematic illustration of a ride-sourcing system.}\label{fig:Figure_Framework}
\end{figure}

The ride-sourcing system can be explained with three key players: traveler (who induce ride requests), service fleet (composed of driver-partners), and service platform (which provides a two-sided market and connects demand and supply). Figure \ref{fig:Figure_Framework} illustrates the interaction of these players in a ride-sourcing market. 
From the demand side, a traveler sends a ride request to the platform with travel details on origin and destination, and the choice of mode (i.e.~single or shared ride). This travel decision is based on the expectation of utility towards the mode of ride-sourcing service which may be affected by socio-demographic characteristics (e.g.~income level, gender, age group, home and work location, car-ownership), travel behavioral parameters (e.g.~alternative specific constants, value-of-time, willingness to pay, mode availability, accessibility to transit service, purpose of trips, location of origin-destination), comparison with other travel alternatives (e.g.~traditional taxi, public transit), previous experiences (historical waiting and travel times, travel comfort), and service fare charged by the system (e.g.~surge pricing, promotion). 

Once the decision has been made by the traveler, the system starts to find an available driver in the service fleet based on the request details that include pick-up and drop-off location and potential revenue. The drivers have significant flexibility in working hour choices and can join and abandon the platform more easily than traditional (full-time) taxi drivers. The drivers' decision to participate can depend on several attributes of operational experience (e.g.~accumulative service time and distance during the shift, the portion of idle time during the shift, remaining fuel for the ride) `shift' behavior (e.g.~shift start and end time, number of shifts), and other effects as well (e.g.~weather conditions).

The common attribute that has a large effect for decisions on both demand and supply is the price. The system estimates and sets the surge multiplier with the aim to balance demand and supply. It generally divides the temporal and spatial domain into discrete intervals (e.g.~10-15min) and cells (e.g.~traffic analysis zone) and determines the imbalance between the number of incoming ride requests and available drivers within same zone during the last time interval. Once a driver decides to participate, they perform a pick-up and drop-off schedule (PUDO) and then get idled after completing a ride service.

\section{Data and Measurement}\label{sec:sec_3_data}
The ride-sourcing data used in our current work includes service details provided by a transportation network company (TNC) over the island of Singapore.
In this study, we focus on weekdays (except public holidays, weekends) in October 2018 for the major ride-sourcing products of a TNC in Singapore. 
In this dataset, detailed service logs at the individual driver level have been collected with more than 7.64 million trip records. Each record includes the following sequence of events: (i) Request sent, (ii) Assignment of vehicle, (iii) Pick-up, and (iv) Drop-off of passenger. To understand the operational characteristics of ride-sourcing, we measured the service performance from the two categories: (a) Operation metrics of service providers and (b) Travel experience of passengers. We characterized the operation metrics of ride-sourcing with the following measurements: service rate, fleet size and utilization, service and operational distance, and surge multiplier. We evaluated the user experience of individual passengers with the waiting time (until pick-up after request sent), travel time and trip speed (during the journey from pick-up to drop-off point). These statistics have been aggregated over the observation period. The summary of experimental setting is in Table \ref{tab:Table_Experimental_Setting}.

\begin{table}[!ht]
	\caption{Experimental setting}\label{tab:Table_Experimental_Setting}
	\begin{center}
		\begin{tabular}{c c | c }
		    \hline
			\multicolumn{2}{l|}{\textbf{Items}} & \textbf{Details}  \\
		    \hline
			\multirow{2}{*}{Data} & Source & Land Transportation Authority (LTA) \\
	        \cline{2-3}
 			& Type &  \makecell{Classified into \texttt{single} and \texttt{shared} by product type of TNC} \\
			\hline
			\multirow{2}{*}{Scope of data} & Coverage & Oct 2018 (23 Weekdays, $P=23$) in Singapore \\
			\cline{2-3}
			& Sample size & \makecell{7.64 million service record\\(6,961,172 (Single) 680,421 (Shared))} \\
			\hline
			\multirow{6}{*}{\makecell{Service \\ criteria}} & \multirow{4}{*}{\makecell{Operation \\ metrics}} & Service rate \\
			& & Fleet size and utilization\\
			& & Service and operational distance \\
			& & Surge multiplier \\
			\cline{2-3}
			& \multirow{2}{*}{\makecell{Travel \\ experience}} & Waiting time \\
			& & In-veh travel-time, Travel distance, Trip speed \\
			\hline
		\end{tabular}
	\end{center}
\end{table}

\section{Empirical Analysis}\label{sec:sec_4_empirical_analysis}
\subsection{Demand Distribution}\label{sec:demand_dist}
The total number of ride requests recorded during this period is around 7.6 million in total (6,963,034 trips for single and 680,805 trips for shared ride); daily ridership averaged around 302k and 29.5k for single and shared rides (the demand pattern varies around 4.5\% with the day of week). Note that the number of canceled requests (by customer or driver) accounts for less than 0.05\% of the total demand.

\begin{figure}[htbp]
  \centering
  \includegraphics[width=0.4\textwidth]{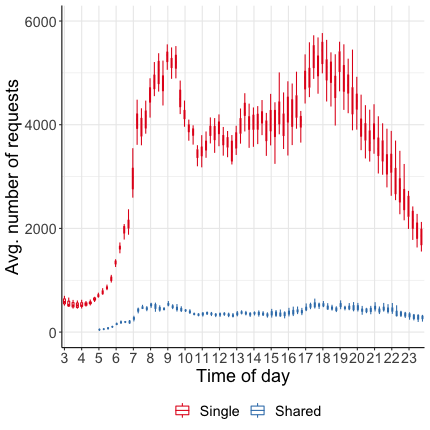}
  \caption{Demand distribution of ride-sourcing by time-of-day.}\label{fig:Figure_Request_Temporal}
\end{figure}

\begin{figure}[htbp]
\centering
\begin{minipage}{0.5\textwidth}
  \centering
    \includegraphics[scale=0.3]{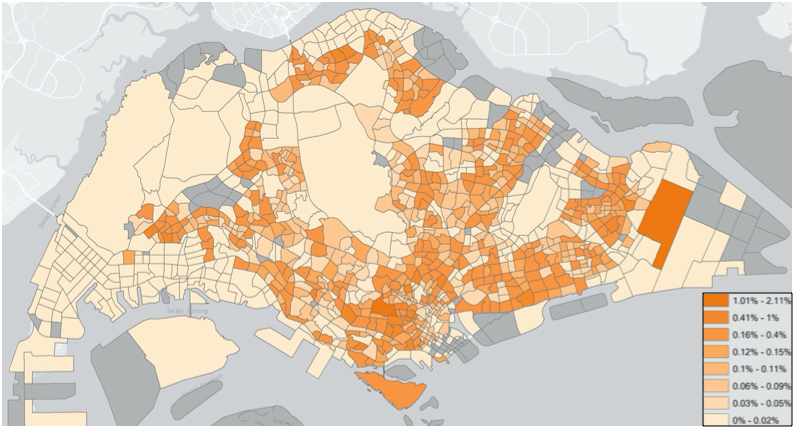}
	\subcaption{Morning peak}\label{fig:Fig_Spatial_Demand_AM}
\end{minipage}
~\hfill
\begin{minipage}{0.5\textwidth}
  \centering
    \includegraphics[scale=0.3]{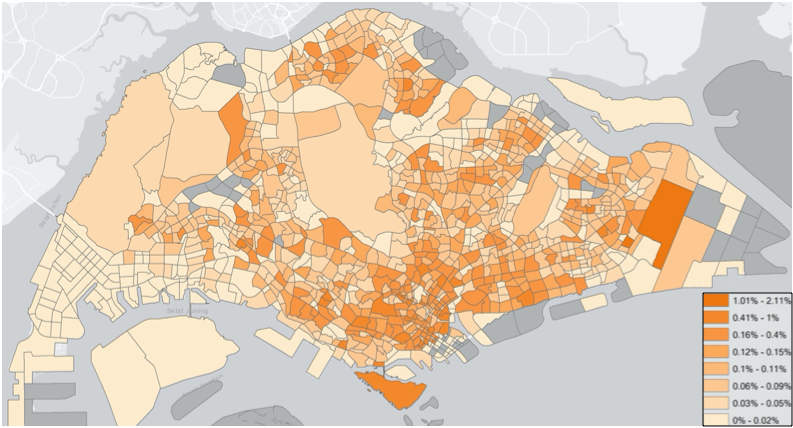}
	\subcaption{Evening peak}\label{fig:Fig_Spatial_Demand_PM}
\end{minipage}
  \caption{Spatial distribution of ride request by zone.}\label{fig:Figure_Request_Spatial}
\end{figure}

Figure~\ref{fig:Figure_Request_Temporal} illustrates the temporal distribution of demand over time-of-day (excluding canceled requests). We have measured the average number of requests throughout the period ($\overline{N}_{Request}$); we found that it increases  significantly during the morning and evening peak with commuting trips, reaching more than 5,000 (service/15min) for single and 530 (service/15min) for shared rides. Large portion of requests came from the residential areas (distributed over the western, eastern, northern part of island) and the central business district (located central/southern part of island) in the morning and evening peak respectively (see Figure \ref{fig:Figure_Request_Spatial}).

\begin{figure}[htbp]
  \centering
  \includegraphics[width=0.8\textwidth]{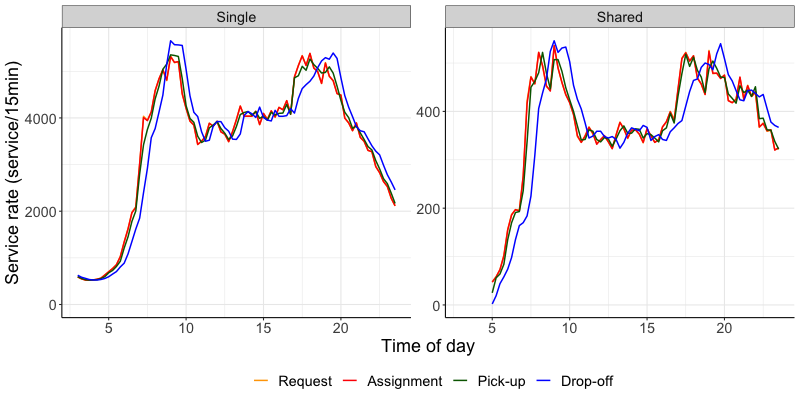}
  \caption{Distribution of average service rate by time-of-day.}\label{fig:Figure_Service_Rate}
\end{figure}

\subsection{Service Metrics}\label{sec:service_metric}
\subsubsection{Service Rate}
Figure \ref{fig:Figure_Service_Rate} shows the distribution of average service rate ($\overline{N}(t)$) which has been averaged throughout the observation period ($p \in P$) by time-of-day ($t \in T$) for each event ($i \in$ \{\textit{Request}, \textit{Assignment}, \textit{Pickup}, \textit{Dropoff}\}): $\overline{N}_{i}(t) = \frac{\sum_{p}^{P} {N_{i}(p,t)}}{P}$.

The request curve (yellow) is equivalent to the average demand distribution shown in Figure~\ref{fig:Figure_Request_Temporal}. Assignment of vehicles (red curve) is performed right after the request is sent, through a booking assignment algorithm \citep{tang2017data}. 
The system needs a search time until the request is matched with an available driver. The average search time has been measured around 20 and 25 sec during the peak for single and shared rides respectively. Also the average (service) queue length has been estimated as 65 (service/15min) for single and 10 (service/15min) for shared rides during the peak periods. The average service rate of assignment curve reaches to around 4,400--5,000 (service/15min) and 450--480 (service/15min) for single/shared ride during the peak periods. From the service log, the request satisfaction rate is nearly 100\% (as the cancellation rate after sending request is minimal, $<0.05\%$), however it should be noted that the `latent' demand, i.e.~would-be passengers who accessed the ride-sourcing app but did not send a request because of low utility on decision factors (such as high fare, long waiting time), has not been captured from the given data.

The horizontal gap between the assignment and pick-up curve implies the waiting time of passengers (after assigned to a driver). Similarly, the in-vehicle travel-time between pick-up and drop-off point can be measured with the gap between green and blue curve. The details on the distribution of waiting and travel time are explained in Section \ref{sec:user_experience}.

\subsubsection{Fleet Size and Utilization}
\begin{figure}[htbp]
\begin{minipage}{0.32\textwidth}
  \centering
    \includegraphics[scale=0.32]{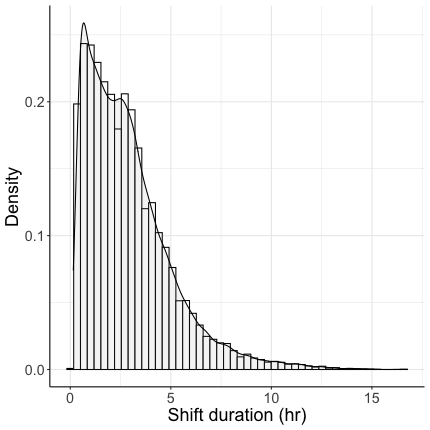}
	\subcaption{Shift duration}\label{fig:Fig_Shift_Duration}
\end{minipage}
~\hfill
\begin{minipage}{0.32\textwidth}
  \centering
    \includegraphics[scale=0.32]{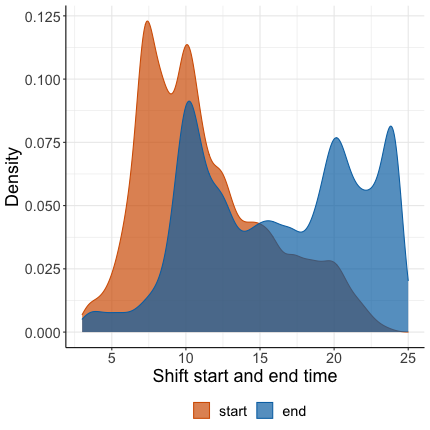}
	\subcaption{Shift start and end time}\label{fig:Fig_Shift_Start_End}
\end{minipage}
~\hfill
\begin{minipage}{0.32\textwidth}
  \centering
    \includegraphics[scale=0.32]{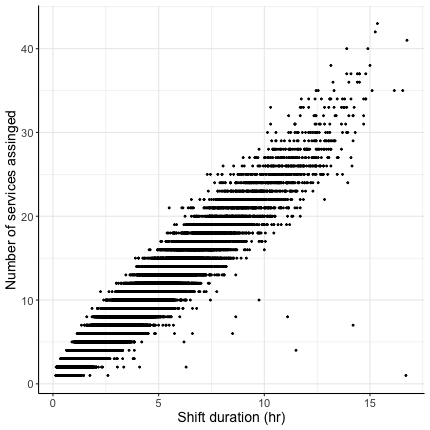}
	\subcaption{Number of services assigned by shift duration}\label{fig:Fig_Shift_vs_service}
\end{minipage}
\caption{Shift behavior of drivers}\label{fig:Figure_5}
\end{figure}

The ride-sourcing system allows drivers to decide to participate in the market anytime, offering to adjust their own flexible schedule from hour to hour in a day. The average fleet size (the total vehicle population which participated in the ride-sourcing service) throughout the month is 25,700 vehicles per day. 
The average number of shifts of a driver during a day is around 2 (varying between 1 to 7 shifts) and the average shift duration is 2.9hr (see Figure \ref{fig:Fig_Shift_Duration}). 
Drivers serve on average 12 trips per shift (Figure \ref{fig:Fig_Shift_vs_service}). Drivers start and end their trips freely during the day, as evidenced in Figure \ref{fig:Fig_Shift_Start_End}: a large portion of drivers start their shift during the morning peak or after lunch time, then end the shift after the end of the morning or evening peak period. 

\begin{figure}[!ht]
  \centering
  \includegraphics[width=0.5\textwidth]{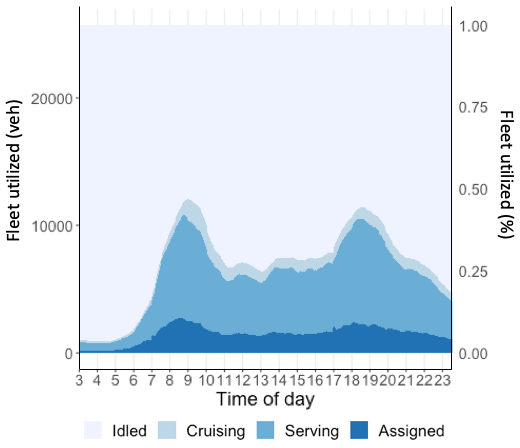}
  \caption{Fleet utilization over time-of-day.}\label{fig:Figure_Fleet_Utilization}
\end{figure}

Figure \ref{fig:Figure_Fleet_Utilization} presents the fleet utilization over time-of-day averaged throughout the period. The set of vehicle status from the vehicle trajectory has been defined as:

\begin{itemize}
    \item \textit{Utilized} if a vehicle is on the way to pick-up point after \textit{assigned} or with passenger(s) by \textit{serving} pick-up and drop-off service.
    \item  \textit{Unutilized} if a vehicle is empty and \textit{cruising} after PUDO until the next service assignment (within 30min) or deactivated in the system and \textit{idled} for significant amount of time (more than 30min).
\end{itemize}

We can see that the utilization rate (\textit{Serving}, \textit{Assigned}) increases to around 50\% during the peak periods. A significant portion of vehicles has been captured during their way to the location of pick-up from current location after got assigned. After \textit{serving} a PUDO, \textit{cruising} vehicles generate empty trips until they are assigned to a next request. Large portion of \textit{idled} vehicles can be explained with the shift behavior of ride-sourcing driver-partners (short shift duration, flexible start and end time, multiple number of shifts); this results in an `effective' fleet size (which drivers are on-line waiting for participating) that is much less than the total fleet size.

\subsubsection{Service and Operating Distance}
The travel distance of service vehicle is key for understanding other metrics such as wait and travel times. For each trip at the driver level, we have the following locations and timestamps:

\begin{itemize}
	\item \textit{Assignment location}: $(\mathbf{x}_{A}, \mathbf{y}_{A}, \mathbf{t}_{A})$ location of the driver reported when the request was \textit{accepted} and \textit{assigned}; the driver has to travel from here to pick up the passenger.
	\item \textit{Pick-up location}: $(\mathbf{x}_{PU}, \mathbf{y}_{PU}, \mathbf{t}_{PU})$ location of the start of the passenger trip.
	\item \textit{Drop-off location}: $(\mathbf{x}_{DO}, \mathbf{y}_{DO}, \mathbf{t}_{DO})$ location of the end of the passenger trip.
\end{itemize}

Based on these coordinates of $\mathbf{x}$ and $\mathbf{t}$, we have distinguished three types of trips of driver: (i) Service trips for \textit{PUDO} of passengers, (ii) \textit{Pick-up (PU)} trips and (iii) \textit{Cruising}. In the PUDO trips, the trajectory is ($\mathbf{x}_{PU}, \mathbf{y}_{PU}) \rightarrow (\mathbf{x}_{DO}, \mathbf{y}_{DO}$), which is straightforward, and the actual distance is available from data set. 

For the case of PU trips, two trajectories can be considered:
\begin{enumerate}
    \item $(\mathbf{x}_{A}, \mathbf{y}_{A}) \rightarrow (\mathbf{x}_{PU}, \mathbf{y}_{PU}))$ ($\mathbf{t}_{A} > \mathbf{t}_{DO}^{prev}$)
    \item $(\mathbf{x}_{DO}^{prev}, \mathbf{y}_{DO}^{prev}) \rightarrow (\mathbf{x}_{PU}, \mathbf{y}_{PU})$ ($\mathbf{t}_{A} < \mathbf{t}_{DO}^{prev}$), where $(\mathbf{x}_{DO}^{prev}, \mathbf{y}_{DO}^{prev}, \mathbf{t}_{DO}^{prev})$ are driver's \textit{previous} drop-off time and location.
\end{enumerate}
We note that this distinction is necessary since drivers can be assigned a next trip even \emph{before} finishing their previous trip. Assigning trips this way allows an operator to offer a more optimal service with less downtime for drivers, especially in busy periods. We measure the distance between $(\mathbf{x}_{A},\mathbf{y}_{A})$ and $(\mathbf{x}_{PU}, \mathbf{y}_{PU})$ for (1), and between $(\mathbf{x}_{DO}^{prev}, \mathbf{y}_{DO}^{prev})$ and $(\mathbf{x}_{PU}, \mathbf{y}_{PU})$ for (2). For shared trips with other passengers, with the exception of the first pick-up, the trajectory is $(\mathbf{x}_{A}, \mathbf{y}_{A}) \rightarrow (\mathbf{x}_{PU}, \mathbf{y}_{PU}) \rightarrow (\mathbf{x}_{DO}, \mathbf{y}_{DO})$, but $(\mathbf{x}_{A}, \mathbf{y}_{A})$ might be a point where a previous passenger is already on board and the distance spent traveling to pick-up ($\mathbf{d_{PU}}$) is part of the trip travel; nevertheless, this can still represent a detour for the first passenger. 

For cruising, we've estimated the distance between the previous drop-off location (the location where the driver finished their previous trip) and the location of the driver reported when the request was accepted (note: this excludes cases when the next trip was already assigned before drop-off). We note that it is unclear what route drivers take between ($\mathbf{x}_{DO}^{prev}, \mathbf{y}_{DO}^{prev}, \mathbf{t}_{DO}^{prev}$) and ($\mathbf{x}_A, \mathbf{y}_A, \mathbf{t}_A$). As defined in the previous section, we consider drivers idled if they are not assigned any trip for a significant amount of time, i.e.~$\mathbf{t}_{A} > (\mathbf{t}_{DO}^{prev} + T_{Idled})$), where $T_I = 30\,\mathrm{min}$; we exclude these drivers from \textit{idled}. 

For each trip, we define the following distances:
\begin{itemize}
	\item $\mathbf{d_{PU}}$, the PU distance between $(\mathbf{x}_{A}, \mathbf{y}_{A})$ and $(\mathbf{x}_{PU}, \mathbf{y}_{PU})$, that is traveled by the driver to meet the passenger. In the case of trips which assigned new service before finishing the previous one ($\mathbf{t}_{A} \leq \mathbf{t}_{DO}^{prev}$), the distance between $(\mathbf{x}_{DO}^{prev}, \mathbf{y}_{DO}^{prev})$ and $(\mathbf{x}_{PU}, \mathbf{y}_{PU})$ is used instead.
	\item $\mathbf{d_{PUDO}}$, the actual PUDO distance with the passenger on-board, the distance between $(\mathbf{x}_{PU}, \mathbf{y}_{PU})$ and $(\mathbf{x}_{DO}, \mathbf{y}_{DO})$.
    \item $\mathbf{d_{C}}$, the cruising distance between from $(\mathbf{x}_{DO}^{prev}, \mathbf{y}_{DO}^{prev})$ to $(\mathbf{x}_{A}, \mathbf{y}_{A})$, i.e.~any travel the driver did between the end of the previous trip and accepting the next one.
\end{itemize}

Note that, for $\mathbf{d_{PUDO}}$, we have the real values reported in the data set. We estimated the other distances ($\mathbf{d_{C}}$, $\mathbf{d_{PU}}$) with the shortest path distance based on the OpenStreetMap road network of Singapore, assuming the drivers are taking the shortest route. We note that this likely underestimates cruising distances, especially if the driver spends a significant time before being assigned to a next trip. The distance variables (average distance per trips and average total traveled distance per day) are summarized in Table \ref{tab:Table_Distance_Measure}.

\begin{table}
	\caption{Summary of operating and service distances}\label{tab:Table_Distance_Measure}
	\centering
	\begin{tabular}{l|l l r r}
		\hline
		\textbf{Category} & \textbf{Measurements} & & \textbf{Avg per trip [km]} & \textbf{Total per day [km]} \\ \hline
		\multirow{3}{*}{\textbf{Operating distance}} & $\mathbf{d_{C}}$ (estimated) & & 1.920 & 320,235 \\
		\cline{2-5}
		& \multirow{2}{*}{$\mathbf{d_{PU}}$ (estimated)} & PU w/o pax & 1.236 & 322,832 \\
		& & PU w/ pax & 1.507 & 10,739 \\
		\hline
		\multirow{2}{*}{\textbf{Service distance}} & \multirow{2}{*}{$\mathbf{d_{PUDO}}$ (real)} & single ride & 11.34 & 3,431,373 \\
		& & shared ride & 13.96 & 413,098 \\ 
		\hline
	\end{tabular}
\end{table}

\subsubsection{Surge Multiplier}
\begin{figure}[htbp]
  \centering
  \includegraphics[width=0.45\textwidth]{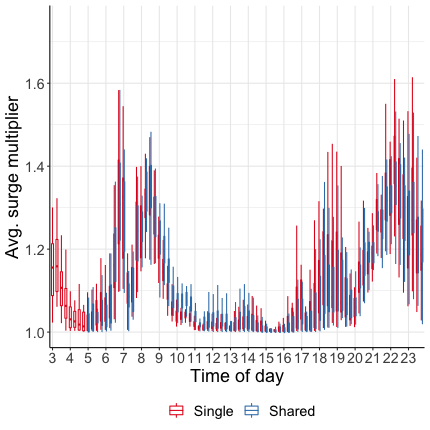}
  \caption{Distribution of surge multiplier by time-of-day.}\label{fig:Figure_Surge_Multiplier}
\end{figure}

Having a surge multiplier encourages idled drivers to participate in the service; it is expected to be employed by the platform when there is an imbalance between demand and supply. The objective in this study is not to reveal the exact underlying mechanism that causes an increase in surge multiplier, but understanding the spatiotemporal variations from the dataset. Figure \ref{fig:Figure_Surge_Multiplier} presents the average surge multiplier over time-of-day, which increased up to 1.6 during peak periods. Table \ref{tab:Table_Surge_Multiplier} lists the average surge multiplier and average fare charged per unit km for each time period. Large surge multipliers have been estimated during the morning peak (1.26-1.28) and off-peak at night (1.33-1.31). This corresponds to the unit fare increasing to around S\$1.8 and S\$1.2 for single and shared rides during the morning and evening peak. 

\begin{table}
    \caption{Average surge multiplier and unit fare by time of day}\label{tab:Table_Surge_Multiplier}
	\centering
	\begin{tabular}{l | c c | c c}
	\hline
		 \multirow{2}{*}{\textbf{Period}} & \multicolumn{2}{c|}{\textbf{Surge multiplier}} & \multicolumn{2}{c}{\textbf{Unit fare (S\$/km)}} \\
		 & Single & Shared & Single & Shared \\
		 \hline
		 \makecell{Morning peak \\ (7am - 9am)} & 1.28 & 1.26 & 1.797 & 1.183 \\
		 \hline
		 \makecell{Day (off-peak) \\ (10am - 4pm)} & 1.02 & 1.03 & 1.580 & 1.092 \\
		 \hline
		 \makecell{Evening peak \\ (5pm - 8pm)}  & 1.14 & 1.11 & 1.807 & 1.211 \\ 
		 \hline
		 \makecell{Night (off-peak) \\ (9pm - 12am)}  & 1.33 & 1.31 & 1.612 & 1.051 \\ 
    \hline
	\end{tabular}
\end{table}

\begin{figure}[htbp]
\begin{minipage}{0.5\textwidth}
  \centering
    \includegraphics[scale=0.25]{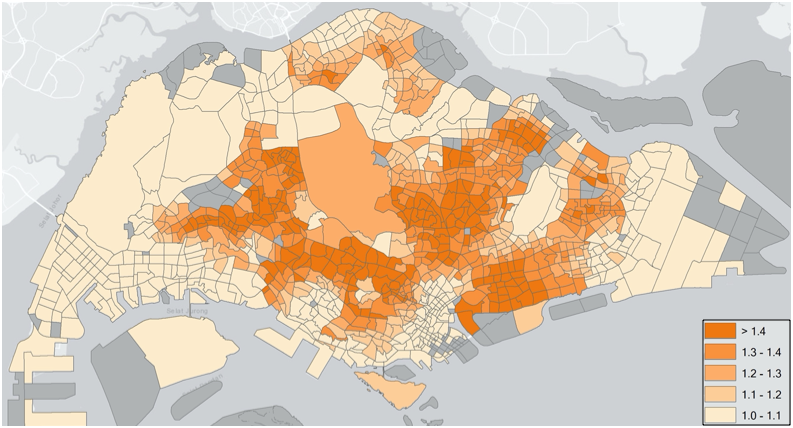}
	\subcaption{Morning peak}\label{fig:Fig_Spatial_Surge_AM}
\end{minipage}
~\hfill
\begin{minipage}{0.5\textwidth}
  \centering
    \includegraphics[scale=0.25]{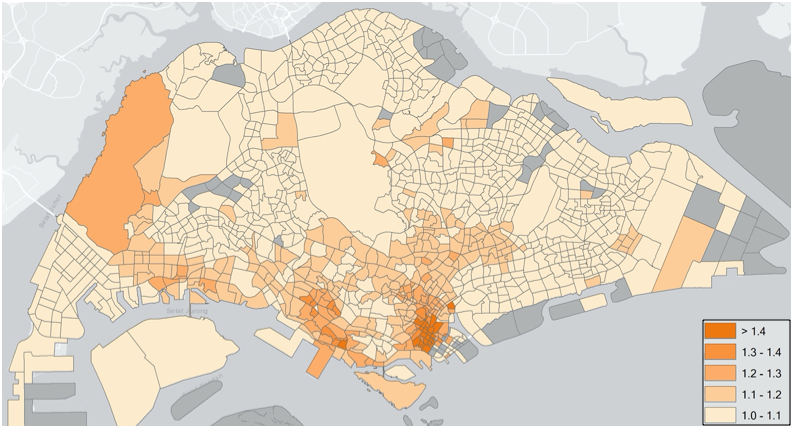}
	\subcaption{Evening peak}\label{fig:Fig_Spatial_Surge_PM}
\end{minipage}
\caption{Spatial distribution of average surge multiplier}\label{fig:Figure_Surge_Multiplier_Spatial}
\end{figure}

The spatial distribution of surge multipliers explains commuting patterns more clearly. During the morning peak, larger surge multipliers can be observed in residential areas in the western and northern parts of Singapore, where commuting trips to work places are generated (Figure \ref{fig:Fig_Spatial_Surge_AM}), while the central business district in the southern area yields larger multipliers (over 1.4) during the evening peak (Figure \ref{fig:Fig_Spatial_Surge_PM}). 

\subsection{User Experience}\label{sec:user_experience}
\subsubsection{Waiting Time}
\begin{figure}[htbp]
  \centering
  \includegraphics[width=0.45\textwidth]{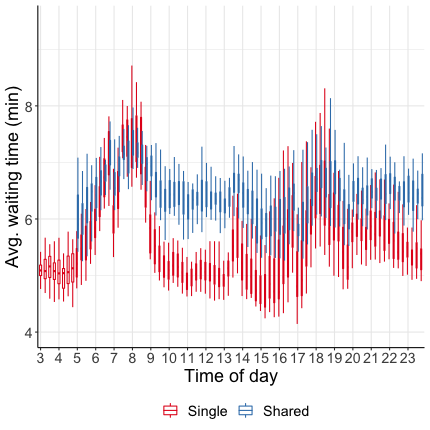}
  \caption{Distribution of average waiting time by time-of-day.}\label{fig:Figure_Waiting_Time}
\end{figure}

The waiting time of travelers has been estimated as the gap between the assignment and pick-up (PU) timestamp as: $\mathbf{wt} = \mathbf{t}_{PU}-\mathbf{t}_{A}$. The average waiting time is shown in Figure \ref{fig:Figure_Waiting_Time} and summarized in Table \ref{tab:Table_Wait_time}. Both curves on single/shared ride increase to 7--8min during the peak periods and reduce to around 5.1min (single) and 6.3min (shared) during  off-peak periods. Overall, the waiting time of shared rides is longer by around 30\% than that of single rides, which can be explained with longer pick-up distances ($\mathbf{d}_{PU}$; refer Section \ref{sec:service_metric} -- again, $\mathbf{d}_{PU}$ is the shortest path estimate between $\mathbf{x}_{A}$ and $\mathbf{x}_{PU}$ on OpenStreetMap).

The averages of $\mathbf{d}_{PU}$ during each period are also summarized in Table \ref{tab:Table_Wait_time} as the distance traveled to pick-up is longer during the morning peak (1.507--1.696km) than the evening peak (1.138--1.55km) and off-peak at night (0.96--1.5km). Also, note that 
each t-test for the distribution of $\mathbf{d}_{PU}$ results in a $p$-value less than 0.01 (close to zero) for each group, indicating that these are statistically significant differences. 

A possible explanation can be found based on the spatial distribution of pick-up locations. Longer $\mathbf{d}_{PU}$ during AM peak can be ascribed to the spatial dispersion of demand as residential areas are spread over the island. During this period, drivers who are located a long distance away from PU points ($\mathbf{x}_{PU}$) might be assigned resulting in longer waiting times (7.08--7.11min on average) even without considering the effect of traffic congestion. Shorter $\mathbf{d}_{PU}$ during PM peak (5pm--8pm) and off-peak at night (9pm--12am) can be explained with the fact that demand departing from work and commercial locations are concentrated in the central business area. Drivers may have learned about the high demand area and prefer to stand by (or cruise) within this area during this period in advance. As a result, we have observed shorter waiting times than in the AM peak during these periods: 5.75--5.95min (single) and 6.50--6.60min (shared). 

\begin{figure}[htbp]
\begin{minipage}{0.5\textwidth}
  \centering
    \includegraphics[scale=0.25]{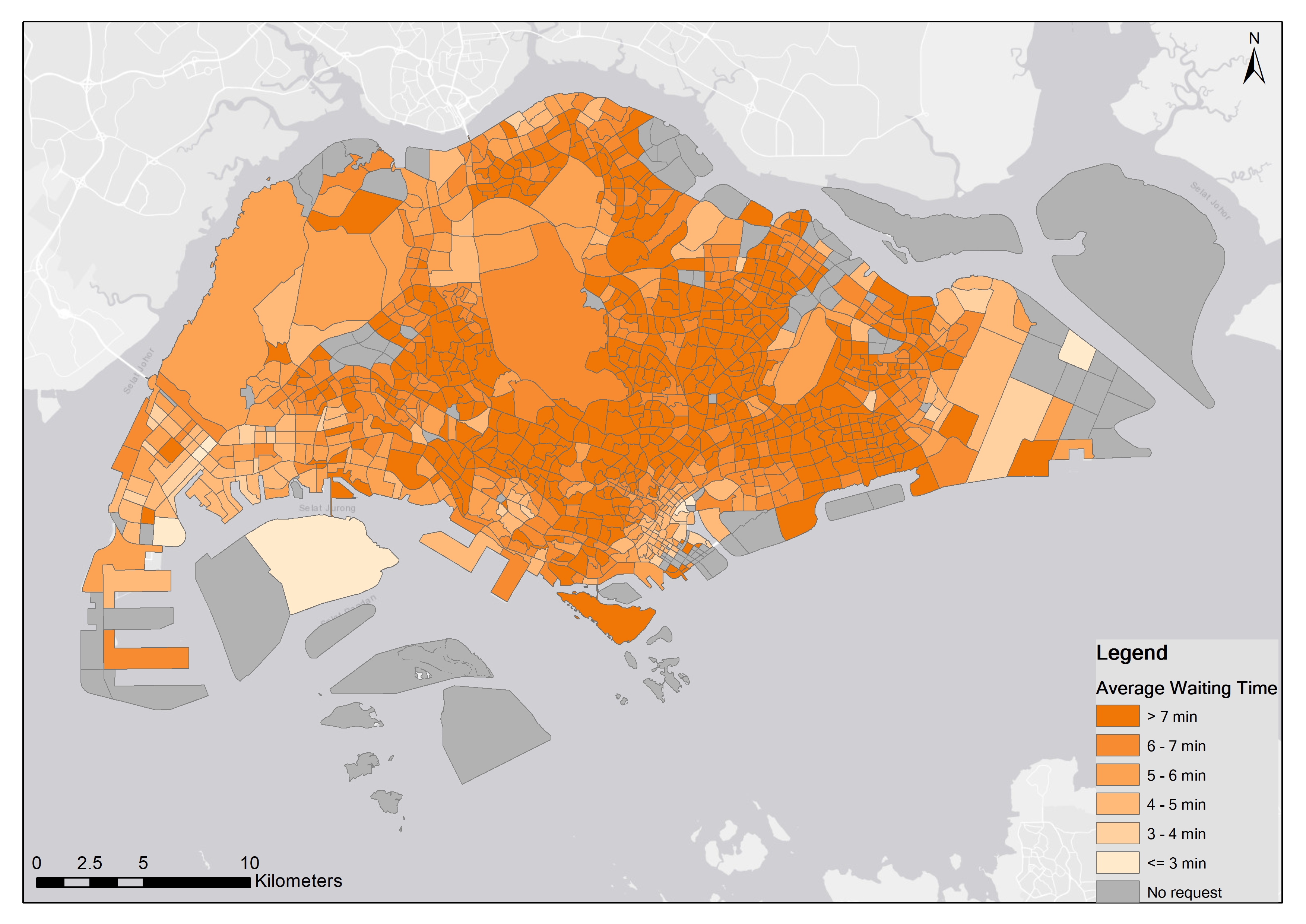}
	\subcaption{Morning peak}\label{fig:wt_am}
\end{minipage}
~\hfill
\begin{minipage}{0.5\textwidth}
  \centering
    \includegraphics[scale=0.25]{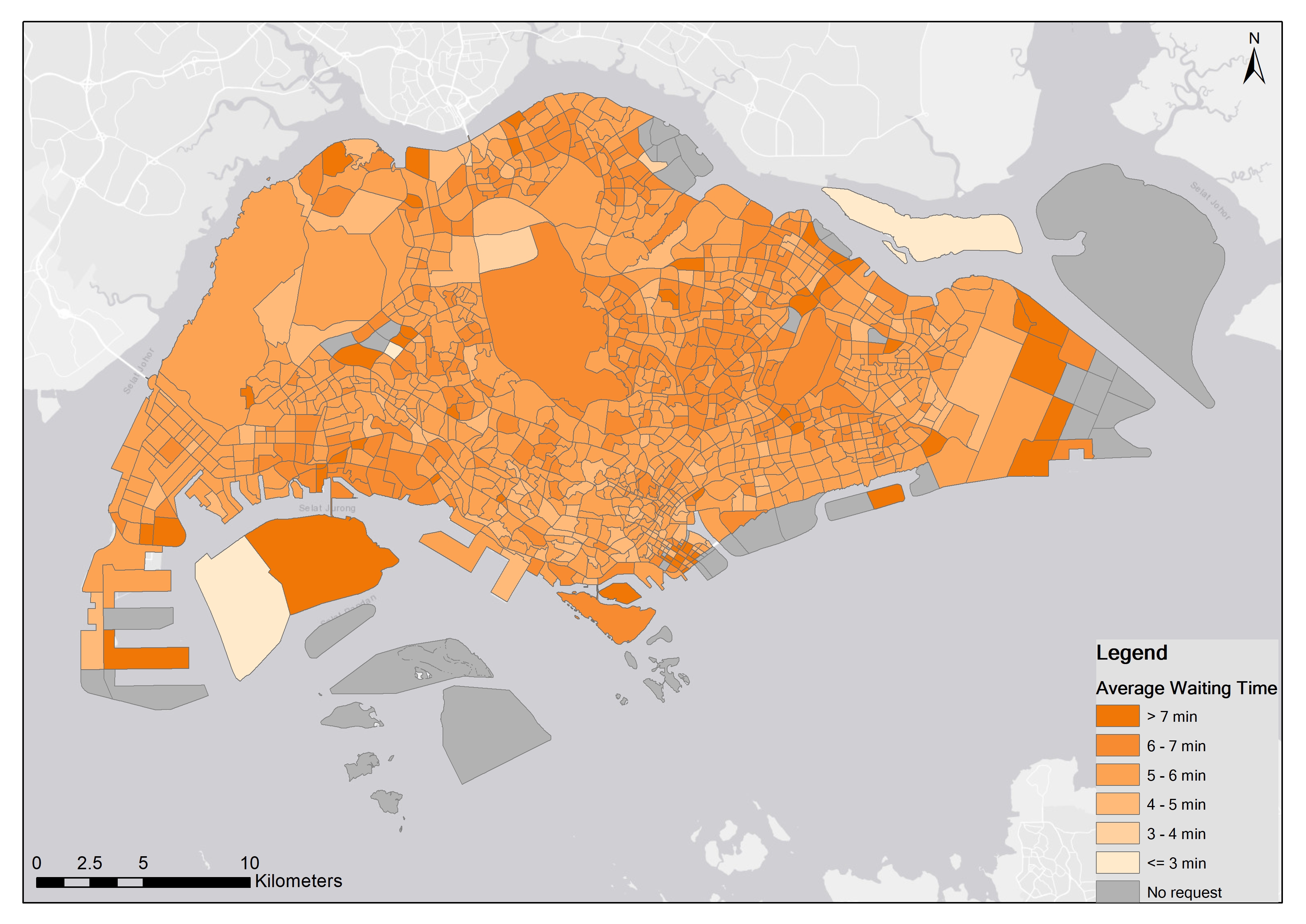}
	\subcaption{Off peak}\label{fig:wt_off}
\end{minipage}
~\hfill
\begin{minipage}{0.5\textwidth}
  \centering
    \includegraphics[scale=0.25]{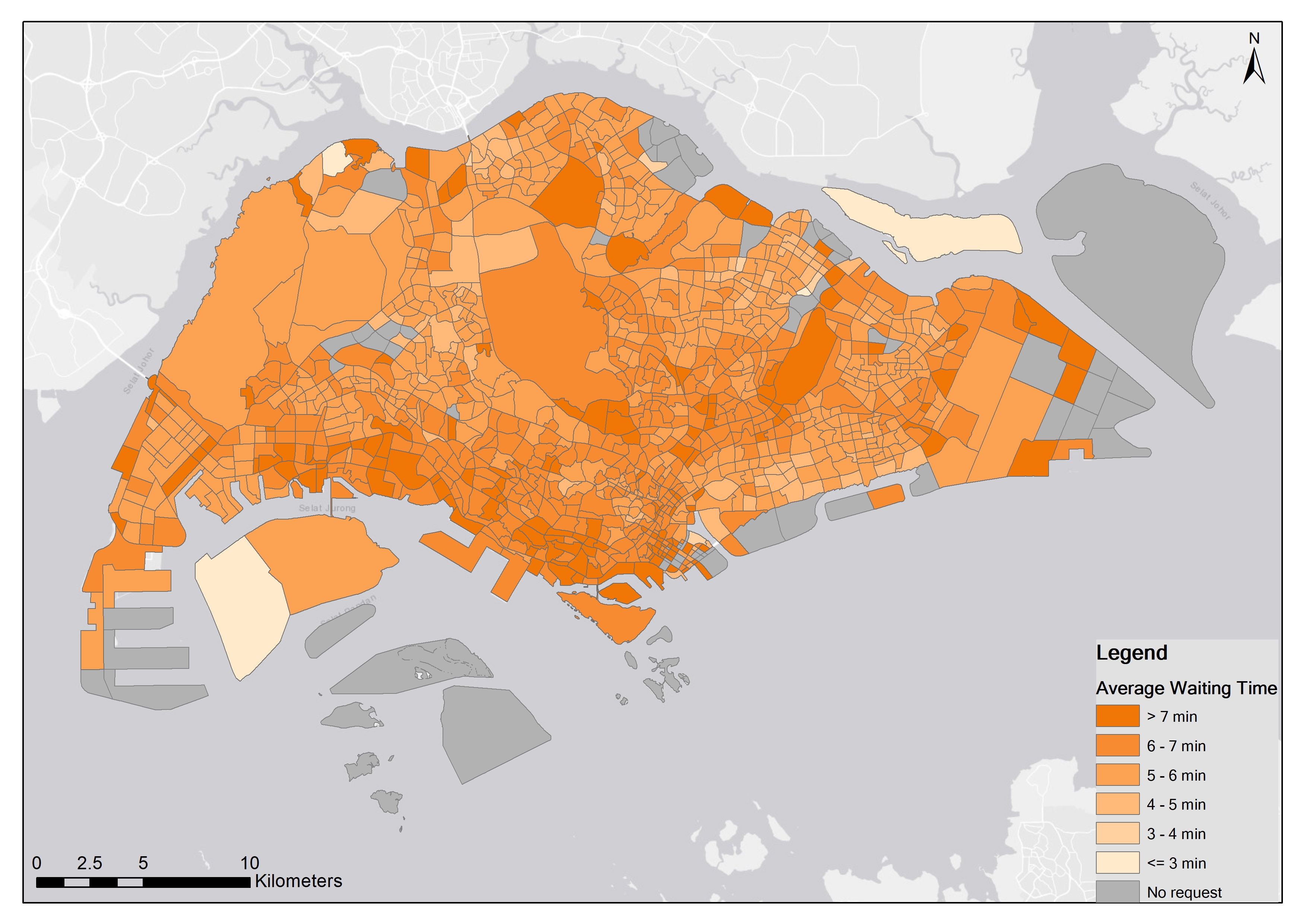}
	\subcaption{Evening peak}\label{fig:wt_pm}
\end{minipage}
\caption{Spatial distribution of average waiting time}\label{fig:Figure_Wait_Time_Spatial}
\end{figure}

Figure \ref{fig:Figure_Wait_Time_Spatial} shows the distribution of average waiting time at each zone over different time periods. As similar with the distribution of demand (in Figure \ref{fig:Figure_Request_Spatial} and Figure \ref{fig:Figure_Surge_Multiplier_Spatial}), longer waiting times have been observed from the residential zones (i.e. relatively remote north western and eastern region from central area) and commercial/business districts (i.e.  industrial area in Jurong west, CBD area at central region of island) during morning and evening peak, respectively.  

\begin{table}
    \caption{Average wait time of passenger and pick-up distance of driver by time of day}\label{tab:Table_Wait_time}
	\centering
	\begin{tabular}{l | c c | c c}
	\hline
		 \multirow{2}{*}{\textbf{Period}} & \multicolumn{2}{c|}{$\mathbf{wt}$ [min]} &  \multicolumn{2}{c}{$\mathbf{d_{PU}}$ [km]} \\
		 & Single & Shared & Single & Shared \\
		 \hline
		 \makecell{Morning peak \\ (7am - 9am)} & 7.08 & 7.11 & 1.507 & 1.696 \\
		 \hline
		 \makecell{Day (off-peak) \\ (10am - 4pm)} & 5.10 & 6.28 & 1.567 & 1.535 \\
		 \hline
		 \makecell{Evening peak \\ (5pm - 8pm)} & 5.95 & 6.49 & 1.138 & 1.550 \\ 
		 \hline
		 \makecell{Night (off-peak) \\ (9pm - 12am)} & 5.75 & 6.60 & 0.959 & 1.500 \\ 
    \hline
	\end{tabular}
\end{table}


\subsubsection{In-vehicle Travel Time, Trip Distance, and Trip Speed}
\begin{figure}[htbp]
\begin{minipage}{0.32\textwidth}
  \centering
    \includegraphics[scale=0.32]{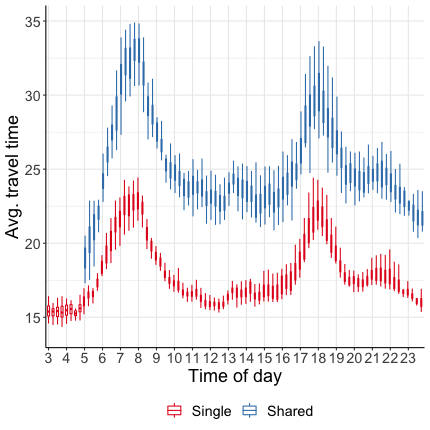}
	\subcaption{Travel time}\label{fig:Fig_Travel_Time}
\end{minipage}
~\hfill
\begin{minipage}{0.32\textwidth}
  \centering
    \includegraphics[scale=0.32]{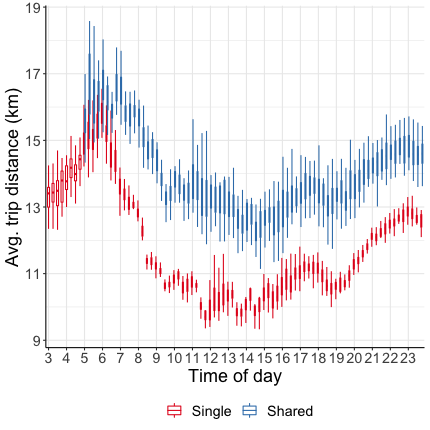}
	\subcaption{Trip distance}\label{fig:Fig_Trip_Dist}
\end{minipage}
~\hfill
\begin{minipage}{0.32\textwidth}
  \centering
    \includegraphics[scale=0.32]{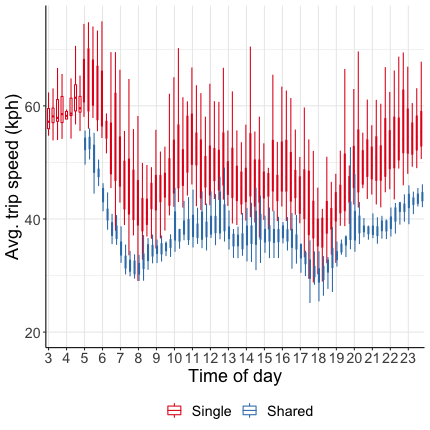}
	\subcaption{Trip speed (weighted by distance)}\label{fig:Fig_Trip_Speed}
\end{minipage}
\caption{Distribution of in-vehicle travel time, distance, and trip speed}\label{fig:Figure_Travel_Time_Dist_Speed}
\end{figure}

This section presents the evolution of user experience over time-of-day with travel-time, trip distance, and trip speed measures. For each trip, the in-vehicle travel-time of the passenger has been measured as: $\mathbf{ivtt}=\mathbf{t}_{DO}-\mathbf{t}_{PU}$. We also estimate the trip speed between the two points as: $\mathbf{v}_{PUDO}=\mathbf{d}_{PUDO}/\mathbf{ivtt}$. Figure \ref{fig:Figure_Travel_Time_Dist_Speed} and Table \ref{tab:Table_Wait_time} summarize the distribution of these variables by time-of-day. Average travel time are longer in shared rides compared to single rides by around 37--45\%. During peak periods, travel times increase to around 21min and 27--32min for single and shared rides respectively. Longer travel times during peak periods can be explained with network congestion. Average travel speed ($\mathbf{v}_{PUDO}$) decreases to around 40--45km/h from 50--60km/h (single) and to 30--35km/h from 40km/h (shared rides) during the peak period. Beside a lower average speed, longer overall travel distances also contribute to longer travel times for shared rides; this also corresponds to experiencing  a `detour' while traveling with other passengers in a vehicle. The travel distance of shared ride is on average 17--24\% longer than that of a single ride.

\begin{table}
    \caption{Average in-vehicle travel-time, distance, and trip speed by time of day}\label{tab:Table_Travel_time}
	\centering
	\begin{tabular}{l | c c | c c | c c}
	\hline
		 \multirow{2}{*}{\textbf{Period}} & \multicolumn{2}{c|}{$\mathbf{ivtt}$ [min]} &  \multicolumn{2}{c|}{$\mathbf{d_{PUDO}}$ [km]} & 
		 \multicolumn{2}{c}{$\mathbf{v_{PUDO}}$ [km/h]} \\
		 & Single & Shared & Single & Shared & Single & Shared \\
		 \hline
		 \makecell{Morning peak \\ (7am - 9am)} & 21.86 & 31.64 & 12.47 & 15.44 & 45.27 & 33.25 \\
		 \hline
		 \makecell{Day (off-peak) \\ (10am - 4pm)} & 16.48 & 23.34 & 10.30 & 13.12 & 50.11 & 40.08 \\
		 \hline
		 \makecell{Evening peak \\ (5pm - 8pm)} & 20.09 & 27.87 & 10.90 & 13.45 & 42.18 & 34.57 \\ 
		 \hline
		 \makecell{Night (off-peak) \\ (9pm - 12am)} & 17.62 & 24.13 & 12.54 & 14.70 & 58.68 & 46.21 \\ 
    \hline
	\end{tabular}
\end{table}


\section{Conclusion}\label{sec:sec_5_conclusion}
This paper empirically analyzed the operational aspects of ride-sourcing using TNC data collected in Singapore. We evaluated the performance based on operation metrics and travel experience. Service demand and supply performance of ride-sourcing showed reproducible patterns over a one month period. Findings suggest important implications in transportation planning, on-demand service operation, and on-demand system modeling. Key highlights can be summarized as follows:

\begin{itemize}
    \item Demand has been characterized with commuting patterns in Singapore. During peak periods, maximum arrival rate of passengers in the system has been estimated more than 5,000 and 530 service/15min for each service type (single and shared). Incoming requests have been assigned to the service fleet with acceptable `departure rate' of 4,400-5,000 service/15min (single) and 450-480 service/15min (shared) and `service delay' of 20-25 sec.
    \item Fleet operational characteristics have been explained with the fleet size and utilization rate over time-of-day. We reported the statistics on drivers' shift behavior with the average number of shifts (varied from 1 to 7), shift duration (2.9h on average), and start and end time of shift (distributed in specific time periods before and after peak periods). These shift variables can be useful for estimating the `effective' fleet size, and also for modeling of driver behavior.
    \item Distribution of surge multiplier showed distinct patterns over time and space, increasing at residential and business/commercial areas for the morning, evening peak, and night time respectively, resulting in an increase in the service fare per unit distance by 14\% compared with off-peak (day time) fare. By merging the demand and fleet operational pattern over spatial zones and time-of-day, there would be a chance to explicitly model the sensitivity for pricing of both travelers and drivers in future studies.
    \item Travel experience with ride-sourcing has been evaluated with waiting time and in-vehicle travel-time distributions. Waiting times increase to around 6--7min during the peak periods, with longer pick-up distance of drivers (in the morning peak) and network congestion (in the evening peak) being significant. Travel time distribution suggests a `detouring' effect of shared rides, which results in longer travel time as well as travel distance. Aggregated waiting and travel times can be used to estimate and calibrate the behavioral models. 
\end{itemize}

Current research stream includes simulating demand-supply interactions of ride-sourcing by measuring `effective' required fleet size and modeling the behavior of travelers and drivers with an emphasis on `shift' behavior and sensitivity for surge pricing. This line of research can be extended further to predict future-looking scenarios such as automated mobility-on-demand and evaluate mobility impacts in multiple dimensions.

\section*{Acknowledgements}
This research is supported by the Singapore Ministry of Transport, Urban Redevelopment Authority, Land Transport Authority, Housing Development Board, Ministry of National Development and the National Research Foundation, Prime Minister’s Office under the Land and Liveability National Innovation Challenge (L2 NIC) Research Programme (L2 NIC Award No. L2NICTDF1-2016-3). Any opinions, findings, and conclusions or recommendations expressed in this material are those of the author(s) and do not reflect the views of the Singapore Ministry of Transport, Urban Redevelopment Authority, Land Transport Authority, Housing Development Board, Ministry of National Development and National Research Foundation, Prime Minister’s Office, Singapore.


\bibliography{References}

\begin{thebibliography}{31}
\expandafter\ifx\csname natexlab\endcsname\relax\def\natexlab#1{#1}\fi
\providecommand{\url}[1]{\texttt{#1}}
\providecommand{\href}[2]{#2}
\providecommand{\path}[1]{#1}
\providecommand{\DOIprefix}{doi:}
\providecommand{\ArXivprefix}{arXiv:}
\providecommand{\URLprefix}{URL: }
\providecommand{\Pubmedprefix}{pmid:}
\providecommand{\doi}[1]{\href{http://dx.doi.org/#1}{\path{#1}}}
\providecommand{\Pubmed}[1]{\href{pmid:#1}{\path{#1}}}
\providecommand{\bibinfo}[2]{#2}
\ifx\xfnm\relax \def\xfnm[#1]{\unskip,\space#1}\fi
\bibitem[{Alonso-Mora et~al.(2017)Alonso-Mora, Samaranayake, Wallar, Frazzoli
  and Rus}]{alonso2017demand}
\bibinfo{author}{Alonso-Mora, J.}, \bibinfo{author}{Samaranayake, S.},
  \bibinfo{author}{Wallar, A.}, \bibinfo{author}{Frazzoli, E.},
  \bibinfo{author}{Rus, D.}, \bibinfo{year}{2017}.
\newblock \bibinfo{title}{On-demand high-capacity ride-sharing via dynamic
  trip-vehicle assignment}.
\newblock \bibinfo{journal}{Proceedings of the National Academy of Sciences}
  \bibinfo{volume}{114}, \bibinfo{pages}{462--467}.
\bibitem[{Battifarano and Qian(2019)}]{battifarano2019predicting}
\bibinfo{author}{Battifarano, M.}, \bibinfo{author}{Qian, Z.S.},
  \bibinfo{year}{2019}.
\newblock \bibinfo{title}{Predicting real-time surge pricing of ride-sourcing
  companies}.
\newblock \bibinfo{journal}{Transportation Research Part C: Emerging
  Technologies} \bibinfo{volume}{107}, \bibinfo{pages}{444--462}.
\bibitem[{Bimpikis et~al.(2019)Bimpikis, Candogan and
  Saban}]{bimpikis2019spatial}
\bibinfo{author}{Bimpikis, K.}, \bibinfo{author}{Candogan, O.},
  \bibinfo{author}{Saban, D.}, \bibinfo{year}{2019}.
\newblock \bibinfo{title}{Spatial pricing in ride-sharing networks}.
\newblock \bibinfo{journal}{Operations Research} \bibinfo{volume}{67},
  \bibinfo{pages}{744--769}.
\bibitem[{Bischoff and Maciejewski(2016)}]{bischoff2016simulation}
\bibinfo{author}{Bischoff, J.}, \bibinfo{author}{Maciejewski, M.},
  \bibinfo{year}{2016}.
\newblock \bibinfo{title}{Simulation of city-wide replacement of private cars
  with autonomous taxis in berlin} .
\bibitem[{Boesch et~al.(2016)Boesch, Ciari and Axhausen}]{boesch2016autonomous}
\bibinfo{author}{Boesch, P.M.}, \bibinfo{author}{Ciari, F.},
  \bibinfo{author}{Axhausen, K.W.}, \bibinfo{year}{2016}.
\newblock \bibinfo{title}{Autonomous vehicle fleet sizes required to serve
  different levels of demand}.
\newblock \bibinfo{journal}{Transportation Research Record}
  \bibinfo{volume}{2542}, \bibinfo{pages}{111--119}.
\bibitem[{Chen and Sheldon(2015)}]{chen2015dynamic}
\bibinfo{author}{Chen, M.K.}, \bibinfo{author}{Sheldon, M.},
  \bibinfo{year}{2015}.
\newblock \bibinfo{title}{Dynamic pricing in a labor market: Surge pricing and
  flexible work on the uber platform}.
\newblock \bibinfo{journal}{UCLA Anderson. URL: https://www. anderson. ucla.
  edu} .
\bibitem[{Cohen et~al.(2016)Cohen, Hahn, Hall, Levitt and
  Metcalfe}]{cohen2016using}
\bibinfo{author}{Cohen, P.}, \bibinfo{author}{Hahn, R.}, \bibinfo{author}{Hall,
  J.}, \bibinfo{author}{Levitt, S.}, \bibinfo{author}{Metcalfe, R.},
  \bibinfo{year}{2016}.
\newblock \bibinfo{title}{Using big data to estimate consumer surplus: The case
  of uber}.
\newblock \bibinfo{type}{Technical Report}. National Bureau of Economic
  Research.
\bibitem[{Comini et~al.(2018)Comini, Erbayat and Pike}]{oecdreport2018}
\bibinfo{author}{Comini, N.}, \bibinfo{author}{Erbayat, B.},
  \bibinfo{author}{Pike, C.}, \bibinfo{year}{2018}.
\newblock \bibinfo{title}{{Taxi, ride-sourcing and ride-sharing services -
  Background Note by the Secretariat}}.
\newblock \bibinfo{type}{Technical Report}.
\newblock \URLprefix \url{OECD Competition Paper}.
\bibitem[{Cramer and Krueger(2016)}]{cramer2016disruptive}
\bibinfo{author}{Cramer, J.}, \bibinfo{author}{Krueger, A.B.},
  \bibinfo{year}{2016}.
\newblock \bibinfo{title}{Disruptive change in the taxi business: The case of
  uber}.
\newblock \bibinfo{journal}{American Economic Review} \bibinfo{volume}{106},
  \bibinfo{pages}{177--82}.
\bibitem[{Fagnant and Kockelman(2014)}]{fagnant2014travel}
\bibinfo{author}{Fagnant, D.J.}, \bibinfo{author}{Kockelman, K.M.},
  \bibinfo{year}{2014}.
\newblock \bibinfo{title}{The travel and environmental implications of shared
  autonomous vehicles, using agent-based model scenarios}.
\newblock \bibinfo{journal}{Transportation Research Part C: Emerging
  Technologies} \bibinfo{volume}{40}, \bibinfo{pages}{1--13}.
\bibitem[{Guda and Subramanian(2017)}]{guda2017strategic}
\bibinfo{author}{Guda, H.}, \bibinfo{author}{Subramanian, U.},
  \bibinfo{year}{2017}.
\newblock \bibinfo{title}{Strategic pricing and forecast communication on
  on-demand service platforms}.
\newblock \bibinfo{journal}{Availbale at SSRN} \bibinfo{volume}{2895227}.
\bibitem[{Guha et~al.(2018)Guha, Demirezen and Kumar}]{guha2018dynamics}
\bibinfo{author}{Guha, S.}, \bibinfo{author}{Demirezen, E.M.},
  \bibinfo{author}{Kumar, S.}, \bibinfo{year}{2018}.
\newblock \bibinfo{title}{Dynamics of competition in on-demand economy: A
  differential games approach}.
\newblock \bibinfo{journal}{Available at SSRN 3263152} .
\bibitem[{Hall et~al.(2015)Hall, Kendrick and Nosko}]{hall2015effects}
\bibinfo{author}{Hall, J.}, \bibinfo{author}{Kendrick, C.},
  \bibinfo{author}{Nosko, C.}, \bibinfo{year}{2015}.
\newblock \bibinfo{title}{The effects of uber’s surge pricing: A case study}.
\newblock \bibinfo{journal}{The University of Chicago Booth School of Business}
  .
\bibitem[{H{\"o}rl et~al.(2019)H{\"o}rl, Ruch, Becker, Frazzoli and
  Axhausen}]{horl2019fleet}
\bibinfo{author}{H{\"o}rl, S.}, \bibinfo{author}{Ruch, C.},
  \bibinfo{author}{Becker, F.}, \bibinfo{author}{Frazzoli, E.},
  \bibinfo{author}{Axhausen, K.W.}, \bibinfo{year}{2019}.
\newblock \bibinfo{title}{Fleet operational policies for automated mobility: A
  simulation assessment for zurich}.
\newblock \bibinfo{journal}{Transportation Research Part C: Emerging
  Technologies} \bibinfo{volume}{102}, \bibinfo{pages}{20--31}.
\bibitem[{Hyland and Mahmassani(2018)}]{hyland2018dynamic}
\bibinfo{author}{Hyland, M.}, \bibinfo{author}{Mahmassani, H.S.},
  \bibinfo{year}{2018}.
\newblock \bibinfo{title}{Dynamic autonomous vehicle fleet operations:
  Optimization-based strategies to assign avs to immediate traveler demand
  requests}.
\newblock \bibinfo{journal}{Transportation Research Part C: Emerging
  Technologies} \bibinfo{volume}{92}, \bibinfo{pages}{278--297}.
\bibitem[{Ke et~al.(2017)Ke, Zheng, Yang and Chen}]{ke2017short}
\bibinfo{author}{Ke, J.}, \bibinfo{author}{Zheng, H.}, \bibinfo{author}{Yang,
  H.}, \bibinfo{author}{Chen, X.M.}, \bibinfo{year}{2017}.
\newblock \bibinfo{title}{Short-term forecasting of passenger demand under
  on-demand ride services: A spatio-temporal deep learning approach}.
\newblock \bibinfo{journal}{Transportation Research Part C: Emerging
  Technologies} \bibinfo{volume}{85}, \bibinfo{pages}{591--608}.
\bibitem[{Oh et~al.(2020a)Oh, Seshadri, Le, Zegras and
  Ben-Akiva}]{oh2020evaluating}
\bibinfo{author}{Oh, S.}, \bibinfo{author}{Seshadri, R.}, \bibinfo{author}{Le,
  D.T.}, \bibinfo{author}{Zegras, P.C.}, \bibinfo{author}{Ben-Akiva, M.E.},
  \bibinfo{year}{2020}a.
\newblock \bibinfo{title}{Evaluating automated demand responsive transit using
  microsimulation}.
\newblock \bibinfo{journal}{IEEE Access} \bibinfo{volume}{8},
  \bibinfo{pages}{82551--82561}.
\bibitem[{Oh et~al.(2020b)Oh, Seshadri, Lima~Azevedo, Kumar, Basak and
  Ben-Akiva}]{oh2020amod_tra}
\bibinfo{author}{Oh, S.}, \bibinfo{author}{Seshadri, R.},
  \bibinfo{author}{Lima~Azevedo, C.}, \bibinfo{author}{Kumar, N.},
  \bibinfo{author}{Basak, K.}, \bibinfo{author}{Ben-Akiva, M.},
  \bibinfo{year}{2020}b.
\newblock \bibinfo{title}{Assessing the impacts of automated mobility-on-demand
  through agent-based simulation: A study of singapore}.
\newblock \bibinfo{journal}{Transportation Research Part A: Policy and
  Practice} \bibinfo{volume}{138}, \bibinfo{pages}{367--388}.
\bibitem[{Rayle et~al.(2016)Rayle, Dai, Chan, Cervero and
  Shaheen}]{rayle2016just}
\bibinfo{author}{Rayle, L.}, \bibinfo{author}{Dai, D.}, \bibinfo{author}{Chan,
  N.}, \bibinfo{author}{Cervero, R.}, \bibinfo{author}{Shaheen, S.},
  \bibinfo{year}{2016}.
\newblock \bibinfo{title}{Just a better taxi? a survey-based comparison of
  taxis, transit, and ridesourcing services in san francisco}.
\newblock \bibinfo{journal}{Transport Policy} \bibinfo{volume}{45},
  \bibinfo{pages}{168--178}.
\bibitem[{Santi et~al.(2014)Santi, Resta, Szell, Sobolevsky, Strogatz and
  Ratti}]{santi2014quantifying}
\bibinfo{author}{Santi, P.}, \bibinfo{author}{Resta, G.},
  \bibinfo{author}{Szell, M.}, \bibinfo{author}{Sobolevsky, S.},
  \bibinfo{author}{Strogatz, S.H.}, \bibinfo{author}{Ratti, C.},
  \bibinfo{year}{2014}.
\newblock \bibinfo{title}{Quantifying the benefits of vehicle pooling with
  shareability networks}.
\newblock \bibinfo{journal}{Proceedings of the National Academy of Sciences}
  \bibinfo{volume}{111}, \bibinfo{pages}{13290--13294}.
\bibitem[{Segui-Gasco et~al.(2019)Segui-Gasco, Ballis, Parisi, Kelsall, North
  and Busquets}]{segui2019simulating}
\bibinfo{author}{Segui-Gasco, P.}, \bibinfo{author}{Ballis, H.},
  \bibinfo{author}{Parisi, V.}, \bibinfo{author}{Kelsall, D.G.},
  \bibinfo{author}{North, R.J.}, \bibinfo{author}{Busquets, D.},
  \bibinfo{year}{2019}.
\newblock \bibinfo{title}{Simulating a rich ride-share mobility service using
  agent-based models}.
\newblock \bibinfo{journal}{Transportation} \bibinfo{volume}{46},
  \bibinfo{pages}{2041--2062}.
\bibitem[{SFMTA(2017)}]{sfmtareport2018}
\bibinfo{author}{SFMTA}, \bibinfo{year}{2017}.
\newblock \bibinfo{title}{2013-2017 Travel Decision Survey Data Analysis and
  Comparison Report}.
\newblock \bibinfo{type}{Technical Report}.
\bibitem[{Statista(2017)}]{statistareport2018}
\bibinfo{author}{Statista}, \bibinfo{year}{2017}.
\newblock \bibinfo{title}{Digital Market Outlook Segment Report}.
\newblock \bibinfo{type}{Technical Report}.
\newblock \URLprefix
  \url{https://www.statista.com/outlook/368/100/ride-hailing-taxi/worldwide}.
\bibitem[{Sun et~al.(2019)Sun, Wang and Wan}]{sun2019model}
\bibinfo{author}{Sun, H.}, \bibinfo{author}{Wang, H.}, \bibinfo{author}{Wan,
  Z.}, \bibinfo{year}{2019}.
\newblock \bibinfo{title}{Model and analysis of labor supply for ride-sharing
  platforms in the presence of sample self-selection and endogeneity}.
\newblock \bibinfo{journal}{Transportation Research Part B: Methodological}
  \bibinfo{volume}{125}, \bibinfo{pages}{76--93}.
\bibitem[{Tang et~al.(2017)Tang, Ow, Chen, Cao, Lye and Pan}]{tang2017data}
\bibinfo{author}{Tang, M.}, \bibinfo{author}{Ow, S.}, \bibinfo{author}{Chen,
  W.}, \bibinfo{author}{Cao, Y.}, \bibinfo{author}{Lye, K.w.},
  \bibinfo{author}{Pan, Y.}, \bibinfo{year}{2017}.
\newblock \bibinfo{title}{The data and science behind grabshare carpooling},
  in: \bibinfo{booktitle}{2017 IEEE International Conference on Data Science
  and Advanced Analytics (DSAA)}, \bibinfo{organization}{IEEE}. pp.
  \bibinfo{pages}{405--411}.
\bibitem[{Vazifeh et~al.(2018)Vazifeh, Santi, Resta, Strogatz and
  Ratti}]{vazifeh2018addressing}
\bibinfo{author}{Vazifeh, M.M.}, \bibinfo{author}{Santi, P.},
  \bibinfo{author}{Resta, G.}, \bibinfo{author}{Strogatz, S.H.},
  \bibinfo{author}{Ratti, C.}, \bibinfo{year}{2018}.
\newblock \bibinfo{title}{Addressing the minimum fleet problem in on-demand
  urban mobility}.
\newblock \bibinfo{journal}{Nature} \bibinfo{volume}{557},
  \bibinfo{pages}{534--538}.
\bibitem[{Wang and Yang(2019)}]{wang2019ridesourcing}
\bibinfo{author}{Wang, H.}, \bibinfo{author}{Yang, H.}, \bibinfo{year}{2019}.
\newblock \bibinfo{title}{Ridesourcing systems: A framework and review}.
\newblock \bibinfo{journal}{Transportation Research Part B: Methodological}
  \bibinfo{volume}{129}, \bibinfo{pages}{122--155}.
\bibitem[{Wang et~al.(2016)Wang, He, Yang and Gao}]{wang2016pricing}
\bibinfo{author}{Wang, X.}, \bibinfo{author}{He, F.}, \bibinfo{author}{Yang,
  H.}, \bibinfo{author}{Gao, H.O.}, \bibinfo{year}{2016}.
\newblock \bibinfo{title}{Pricing strategies for a taxi-hailing platform}.
\newblock \bibinfo{journal}{Transportation Research Part E: Logistics and
  Transportation Review} \bibinfo{volume}{93}, \bibinfo{pages}{212--231}.
\bibitem[{Yang and Yang(2011)}]{yang2011equilibrium}
\bibinfo{author}{Yang, H.}, \bibinfo{author}{Yang, T.}, \bibinfo{year}{2011}.
\newblock \bibinfo{title}{Equilibrium properties of taxi markets with search
  frictions}.
\newblock \bibinfo{journal}{Transportation Research Part B: Methodological}
  \bibinfo{volume}{45}, \bibinfo{pages}{696--713}.
\bibitem[{Zha et~al.(2016)Zha, Yin and Yang}]{zha2016economic}
\bibinfo{author}{Zha, L.}, \bibinfo{author}{Yin, Y.}, \bibinfo{author}{Yang,
  H.}, \bibinfo{year}{2016}.
\newblock \bibinfo{title}{Economic analysis of ride-sourcing markets}.
\newblock \bibinfo{journal}{Transportation Research Part C: Emerging
  Technologies} \bibinfo{volume}{71}, \bibinfo{pages}{249--266}.
\bibitem[{Zhang and Pavone(2016)}]{zhang2016control}
\bibinfo{author}{Zhang, R.}, \bibinfo{author}{Pavone, M.},
  \bibinfo{year}{2016}.
\newblock \bibinfo{title}{Control of robotic mobility-on-demand systems: a
  queueing-theoretical perspective}.
\newblock \bibinfo{journal}{The International Journal of Robotics Research}
  \bibinfo{volume}{35}, \bibinfo{pages}{186--203}.

\end{thebibliography}

\end{document}